\documentclass[%
 aip,
 amsmath,amssymb,
preprint,%
]{revtex4-1}

\usepackage{graphicx}
\usepackage{dcolumn}
\usepackage{bm}

\usepackage[utf8]{inputenc}
\usepackage[T1]{fontenc}
\usepackage{mathptmx}
\usepackage{etoolbox}

\newcommand{\nid}{\noindent}
\usepackage[version=3]{mhchem}
\usepackage{lipsum}

\makeatletter
\def\@email#1#2{%
 \endgroup
 \patchcmd{\titleblock@produce}
  {\frontmatter@RRAPformat}
  {\frontmatter@RRAPformat{\produce@RRAP{*#1\href{mailto:#2}{#2}}}\frontmatter@RRAPformat}
  {}{}
}%
\makeatother
\begin{document}

\preprint{AIP/123-QED}

\title{Entropy governs the structure and reactivity of water dissociation under electric fields}
\author{Yair Litman}
 \affiliation{Yusuf Hamied Department of Chemistry,  University of Cambridge,  Lensfield Road,  Cambridge,  CB2 1EW, UK}
 \affiliation{Max
Planck Institute for Polymer Research, Ackermannweg 10, 55128 Mainz,
Germany}
 \email{litmany@mpip-mainz.mpg.de}
  \email{ am452@cam.ac.uk}
\author{Angelos Michaelides}
 \affiliation{Yusuf Hamied Department of Chemistry,  University of Cambridge,  Lensfield Road,  Cambridge,  CB2 1EW, UK}

\date{\today}

\begin{abstract}
The response of water to electric fields is critical to the performance and stability of electrochemical devices, and the selectivity of enzymatic, atmospheric, and organic reactions. A key process in this context is the water (auto)dissociation reaction (WD), which governs acid-base aqueous chemistry and shapes reaction rates and mechanisms. Despite its significance, the thermodynamics of the WD reaction in electrified environments remains poorly understood. Here, we investigate the WD reaction under external electric fields using \textit{ ab initio} molecular dynamics simulations within the framework of the modern theory of polarization. Our results reveal that strong electric fields dramatically enhance the WD reaction, increasing the equilibrium constant by several orders of magnitude. 
Moreover, we show that the applied field transforms the WD reaction from an entropically hindered process to
an entropy-driven one. 
Analysis shows that this is because the electric field alters the tendency of ions to be structure makers or structure breakers. 
By highlighting how strong electric fields reshape solvent organization and reactivity, this work opens new avenues for designing aqueous electro-catalysts that leverage solvent entropy to enhance their performance.
\end{abstract}

\maketitle

The catalytic effect of electric fields in chemical reactions is a well-established and ubiquitous phenomenon. Many enzymes rely on localized fields to lower activation barriers~\cite{Warshel_ChemRev_2006,Fried_AnnRev_2017,Saura_PNAS_2022},  electrochemical redox processes fundamentally depend on electric-field-driven charge transfer~\cite{Bard_book}, and even non-redox organic reactions can be substantially accelerated in the presence of oriented-external electric fields~\cite{Shaik_NatChem_2016,Aragones_Nature_2016,Joy_JACS_2020}.

In aqueous environments, large electric fields can arise naturally due to molecular fluctuations, hydrogen-bond dynamics, and charge transfer processes~\cite{Flor_Science_Laage,Geissler_Science_2001,Reischl_MolPhys_2009}. These fluctuations are believed to be enhanced in anisotropic environments, such as interfaces, with estimated intensities reaching up to 0.1 V/\AA, even though the associated average field magnitudes remain significantly smaller~\cite{Hao_NatComm_2022}.
The extent to which naturally occurring field fluctuations are large enough to drive chemical reactions is an active area of research, and while this link has been falsified in some cases~\cite{Eatoo_ChemSci_2024}, it remains an open hypothesis invoked to rationalize the phenomenon of "on-water catalysis"~\cite{Ruiz-Lopez_NatRev_2020}. 

An elemental reaction sensitive to electric fields is the water (auto)dissociation (WD) reaction, 2 H$_2$O $\rightarrow$ H$_3$O$^+$ + OH$^-$. This reaction is fundamental to aqueous chemistry, forming the basis of all acid–base equilibria. While the WD reaction is well studied under standard conditions \cite{Grifoni_PNAS_2019,Dasgupta_JPCL_2025,Joutsuka_JPCB_2022}, its behavior under strong electric fields remains far less explored~\cite{Rothfuss_JElectroChem_2003,Zhou_Nature_2018,Pinkerton_Langmuir_1999,Saitta_PRL_2012,Cassone_JPCL_2020,Martins-Costa_Angewandte_2025}. In a recent study, Cai \textit{et al.} \cite{Cai_NatCom_2022} studied the graphene–water interface and elegantly correlated the interfacial electric fields with WD reaction rates. They observed pronounced rate enhancements at fields exceeding 0.01 V/\AA, consistent with earlier \textit{ab initio} simulations in bulk water that identified a dissociation threshold near 0.3 V/Å~\cite{Saitta_PRL_2012}. The acceleration of water dissociation rates with increasing electric field strength is intuitive: electric fields stabilize the separation of charged fragments that lead to larger dipole moments. Both experiments and simulations consistently confirm this trend. 
However, a systematic investigation of the thermodynamics of this process under electric fields has remained elusive.

Understanding the WD reaction at finite electric fields is a fundamental prerequisite to elucidating more complex reactions, since small field-induced shifts in the autoionization equilibrium can lead to large changes in local H$^+$ or OH$^-$ concentrations, substantially altering reaction rates and mechanisms. By studying the WD reaction in bulk, where interfacial effects are absent by design, one could establish an important baseline against which more complex interfacial phenomena can be rigorously assessed. In this work, we conduct extensive \textit{ab initio} molecular dynamics (AIMD) simulations to probe the thermodynamics of the WD reaction across a range of temperatures and applied field strengths. 
By quantifying the reaction energy ($\Delta U$) and entropy ($\Delta S$) as functions of the external field, we find that $\Delta U$ remains large and positive with only a minimal dependence on field strength, contrary to common expectations. Strikingly, the reaction shifts from being entropically hindered at zero field to entropically driven at a field of 0.36 V/\AA. We attribute this behavior to structural changes in the hydrogen-bond network, arising from the interplay between field-induced ordering and ion-mediated disruption. Entropic contributions are often overlooked in computational electrocatalysis studies that rely on static calculations to estimate reaction energetics under bias~\cite{CHE,Ringe_Chem_review}.  Our findings show that neglecting entropic changes can lead to qualitatively incorrect conclusions when studying aqueous reactions.  At the same time, entropy could serve as a previously underexplored dimension for understanding aqueous reactivity in strong electric fields and for guiding the design and optimization of aqueous electrochemical systems.

\begin{figure}
\centering
\includegraphics[width=0.5\textwidth]{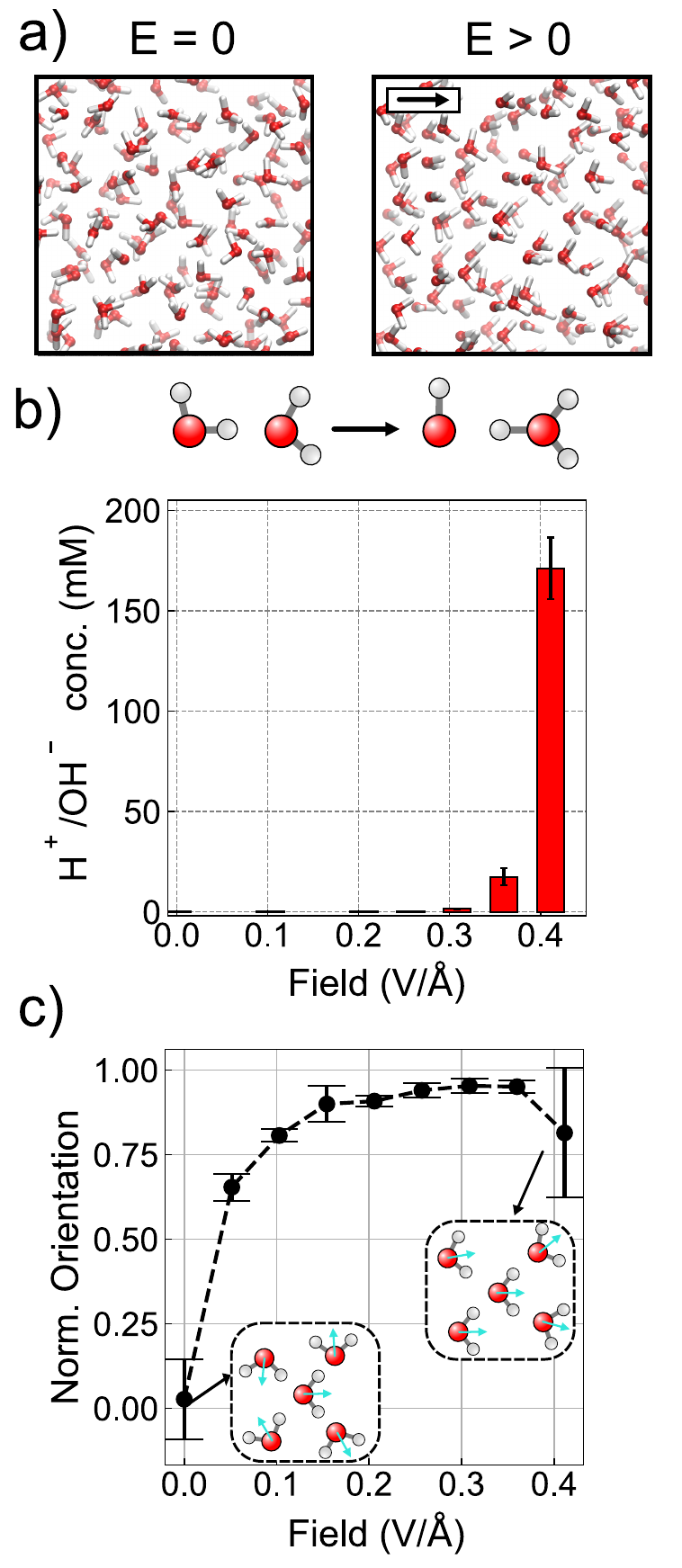}
\caption{\textbf{Reactivity and structuring of water under electric fields.}  a) Schematic representation of the model system in the absence (left) and presence (right) of an external electric field. The external electric field is coupled to the system through the electric enthalpy functional (see main text) and leads to a net polarization. The arrow in the right panel inset depicts the field direction.
b) Proton defect concentration as a function of the field strength. The threshold for the detection of WD reaction events in the employed setup is $\textbf{E}=0.3 $ V/\AA.
c) Average orientation of molecular dipoles along the field direction for pure water.  A value of 1.00 represents a perfect alignment between the molecular dipoles and the external field.
}
\label{fig:sketch}
\end{figure}

\subsection*{External electric fields reduce the water pKw}

Finite electric fields in periodic density functional theory (DFT) can be treated within the Modern Theory of Polarization \cite{Resta_RevModPhys_1994,Resta_PRL_1998,King_PRB_1993,SPALDIN_JSolidStateCHem_2012}. Within this framework, the system's response to an external field, $\textbf{E}$, is described by the electric enthalpy functional \cite{Umari_PRL_2002,Souza_PRL_2002,Stengel_NatPhys_2009} defined as 

        \begin{equation}
       F(\nu,\textbf{E}) = E_\text{KS}(\nu) - \Omega  \textbf{E} \cdot \textbf{P},
       \end{equation}
where $\nu $ represents the ionic and electronic degrees of freedom, $E_\text{KS}$ is the Kohn–Sham energy, $\Omega$ the cell volume, and $\textbf{P}$ the Berry-phase
polarization \cite{Berry_1984,Resta_RevModPhys_1994,Resta_PRL_1998}. 
We performed AIMD based on DFT for bulk liquid water in the NVT ensemble at 330 K using the revPBE exchange-correlation functional~\cite{revPBE} augmented with  D3 dispersion corrections \cite{D3}(Fig. \ref{fig:sketch}a). Simulations were carried out across a range of different field strengths, up to \textbf{E} = 0.4 V/\AA. Further methodological details, including sensitivity tests with respect to the exchange-correlation functional and system size, are provided in the Methods section and Supporting Information (SI).

Fig. \ref{fig:sketch}b shows the equilibirum proton and hydroxide concentration for the different simulations.
In agreement with previous studies \cite{Saitta_PRL_2012,Cassone_JPCL_2020}, we observe WD events above a threshold of \textbf{E} =0.36 V/\AA, above which, the number of ions dramatically increases, reaching a value of 170 $\pm$ 15 mM at  \textbf{E} =0.41 V/\AA. These concentrations represent time-averaged values, as the proton and hydroxide species are short-lived, with sub-picosecond lifetimes before undergoing rapid recombination
 \cite{Hassanali_PNAS_2011}.

The autoionization constant, pKw, can be readily computed at 
a field strength of $\textbf{E} =0.36$ V/\AA~ or higher from the average ion concentrations of proton and hydroxide ions. However, below this threshold,
dissociation events are not observed during the timescales of our simulations, necessitating the use of enhanced sampling techniques to drive the reaction.
Several collective variables and advanced sampling techniques have been developed to address this challenge \cite{Sprik_ChemPhys_2000,Grifoni_PNAS_2019, DeLaPuente_JACS_2023,Dasgupta_JPCL_2025,Joutsuka_JPCB_2022,Liu_PRL_2023,Andrade_PNAS_2023}. 
In this work, we employed the umbrella integration scheme \cite{Kastner_JCP_2005,Kastner_JCP_2006} with  
a single collective variable constructed from the coordination number, $n_\text{cov}$, of a selected water molecule that smoothly transitions from reactant ($n_\text{cov}\sim2$) to products ($n_\text{cov}\sim1$) representing intact water and hydroxide species, respectively.  \cite{Sprik_ChemPhys_2000,Joutsuka_JPCB_2022}.
The first two columns of Table \ref{tab:Free_energy} provide the values of pKw at selected field strengths.
In the absence of an external electric field, our simulations predict a value of 14.7 $\pm$ 0.3, in good agreement with experiments~\cite{Holzapfel_JCP_1969}. As the field increases,  the pKw decreases monotonically, reaching a value of 2.2 $\pm$ 0.2 at $\textbf{E}=0.41$ V/\AA~, the highest considered field strength.

In the absence of an external field, the orientations of water molecules yield zero net polarization (see Fig  \ref{fig:sketch}c). Once the field is applied, the molecular dipoles progressively align with the field direction, with the degree of alignment increasing with field strength. This trend continues until dielectric saturation occurs at $\textbf{E} = \sim 0.25$ V/\AA, where the average reorientation reaches a plateau value.
The decrease in pKw becomes more pronounced above approximately $\textbf{E}=0.2$ V/\AA, coinciding with the change in orientational behavior of water molecules shown in Fig.  \ref{fig:sketch}c. A correlation between enhanced WD and molecular orientation indicates that water molecules in a highly ordered hydrogen-bond network become more susceptible to external electric fields, pointing out a possible important entropic contribution to the thermodynamics of  WD under strong fields.

\begin{table}
\centering
\begin{tabular}{|c|c|c|c|c|}
\hline
$\textbf{E}$ (V/\AA)& pKw &$\Delta F$ (kJ/mol) & $\Delta U$ (kJ/mol)& $\Delta S$ (J/K mol) \\ \hline
0.00 (Exp.) & 14.3 $\pm$ 0.1  & 82.5 $\pm$ 0.1    & 59.5 $\pm$ 0.1 &  -77.2 $\pm$ 0.5  \\ 
\hline
0.00 &  14.7 $\pm$ 0.3 &  84.5 $\pm$ 1.5  & 53.0 $\pm$ 10.3 &  -96.1 $\pm$ 29.8  \\
0.18 &  13.2 $\pm$ 0.3 &  75.8 $\pm$ 1.5  & 80.5 $\pm$ 10.3 &   13.5 $\pm$ 29.8  \\
0.36 &  4.6  $\pm$ 0.3 &  26.6 $\pm$ 1.0  & 56.2 $\pm$ 6.3  &   89.7 $\pm$ 18.1  \\
0.41 &  2.2  $\pm$ 0.2 &  13.7 $\pm$ 0.5  & -  &  - \\
\hline
\end{tabular}

\caption{pKw and free energy decomposition of WD reaction at different external electric field strengths. Experimental values are extracted from Ref. \cite{Holzapfel_JCP_1969}. 
pKw values are computed as, $\mathrm{pKw}=\Delta F/(RT ln (10))$ where R is the gas constant, and T is the temperature, 330 K.}
\label{tab:Free_energy}
\end{table}

\section*{Field-induced WD is entropically driven}

\begin{figure*}
\centering
\includegraphics[width=0.9\columnwidth]{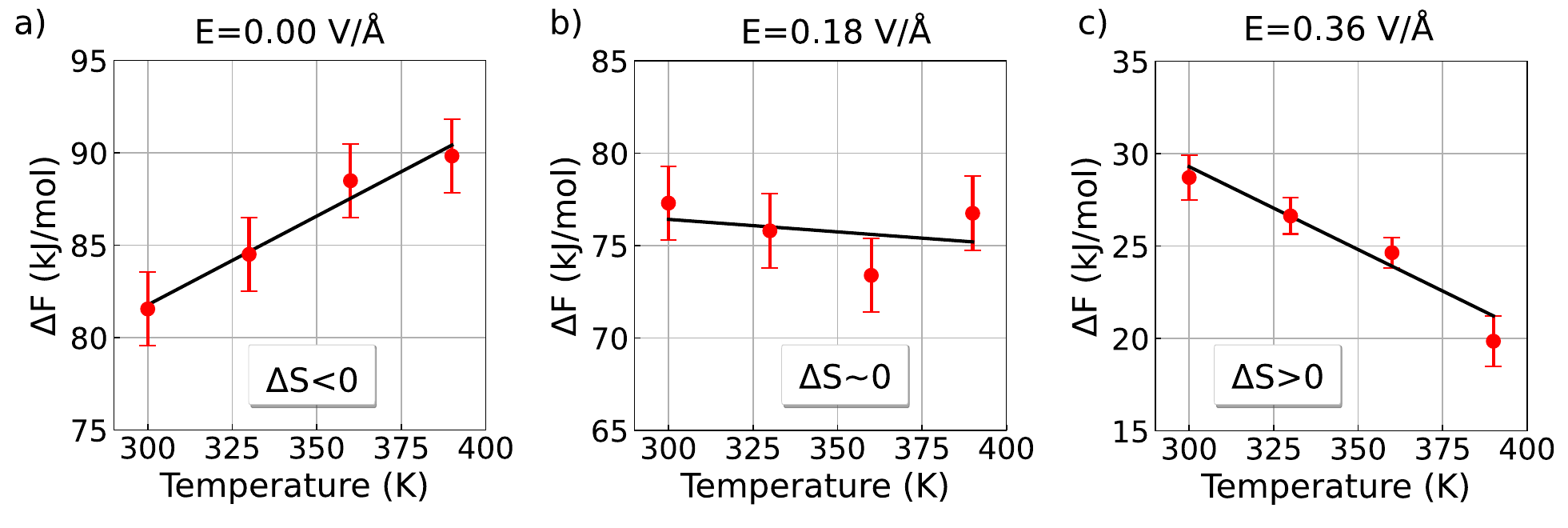}
\caption{\textbf{Dramatic change of temperature dependence of the WD reaction free energy under electric fields.} Helmholtz free energy associated with the WD reaction as a function of temperature and selected values of electric field strengths.}
\label{fig:Free_energy}
\end{figure*}

To disentangle the entropic and energetic contributions of the WD reaction under electric fields, we performed simulations at multiple temperatures.
Fig. \ref{fig:Free_energy}a shows the computed Helmholtz free energy between 300 K and 390 K at three different field strengths.
From these data, we extracted the reaction energy ($\Delta U $) and reaction entropy  ($\Delta S$) by fitting to $\Delta F = \Delta U - T \Delta S$. The results are summarized in Table \ref{tab:Free_energy}.
In the absence of electric fields, our simulations reproduce the experimental value, showing a large energetic cost ($\Delta U$= 53.0 $\pm$ 10.3 kJ/ mol)
and substantial  entropic penalty ($\Delta S$=-96.1 $\pm$ 29.8 J/K mol)  

At finite electric fields, while $\Delta U$ remains large and positive, there is a dramatic increase of  $\Delta S$ by almost 200 J/K mol. 
To confirm these surprising results, we repeated the finite field simulation with another exchange-correlation functional, namely BLYP, and confirmed the presence of a large positive entropic contribution to the reaction free energy (see Fig. S1 in the SI). In the following sections, we examine the structural changes that help to rationalize this unexpected behavior.

\subsection*{Weak and moderate electric fields strengthen the hydrogen-bond network }

We begin by analyzing the field strengths below $\textbf{E} = 0.40 $ V/\AA, where water dissociation remains a minor process.
In Fig. \ref{fig:structure}a-c, we present the O-O and O-H radial distribution functions. The application of an electric field shortens the O-O distance of neighboring waters (first peak of the gr$_\text{OO}$),  elongates the covalent intramolecular O-H bonds (first peak of the gr$_\text{OH}$),  and shortens the 
intermolecular H$\cdots$O distances (second peak of the gr$_\text{OH}$). These are all clear indications of the strengthening of the hydrogen-bond network with applied fields. 
This effect is further accompanied by the field-induced red shift of the O-H stretching vibrations, as shown in panel d~\cite{Cassone_PCCP_2019,Joll_NatCom_2024}.

To characterize the local angular order
of the water molecules, we used the tetrahedral order parameter, $Q$
\cite{CHAU_MOlPhys_1998,
Dubou_JPCB_2015}. In its scaled form~\cite{Errington_Nature_2001}, $Q$ varies from 0 for an ideal gas to 1 for a perfect tetrahedron (see definition in the Methods section).  In Fig. \ref{fig:structure}e, we present the corresponding $Q$  distributions obtained at various fields. In the absence of an external field, we observe a bimodal distribution 
consistent with previous simulations of water \cite{Errington_Nature_2001,Dasgupta_JPCL_2025}. As the field increases, the distribution shifts to the right, with the high-$Q$ peak increasing at the expense
of the low-$Q$ shoulder, leading to a quasi-isosbestic point at $Q=0.74$.
This trend reflects an increasing population of tetrahedrally coordinated water molecules, similar to structural changes observed upon cooling~\cite{Paolantoni_JPCA_2009,Morawietz_JPCL_2018}.

Water molecules can be classified by the number of hydrogen-bonds they donate (Don) and accept (Acep). For example, a water molecule that donates one hydrogen-bond and accepts two is labeled as Don-2Acep. In Fig. \ref{fig:structure}f,  we show the evolution of the 2Don-2Acep, 2Don-1Acep, 1Don-2Acep, and 1Don-1Acep populations as a function of the field strength. Consistent with the behavior of $Q$, the 2Don-2Acep population increases from 0.50 at $\textbf{E}= 0.00$ V/\AA~ to 0.84 at $\textbf{E}=0.36$ V/\AA. Thus, the electric field induces a continuous conversion of disordered configurations into more tetrahedral, ice-like structures. Such an 'electrofreezing' effect has been proposed in theoretical studies \cite{Cassone_NatCom_2024}, although its experimental verification remains elusive \cite{Peleg_JCPCC_2019}.

The formation of new hydrogen-bonds or the strengthening of existing ones leads to energy and mobility changes that directly translate into negative formation enthalpy and entropy.  In our simulations, this value can be estimated from the temperature dependence of the equilibrium constant for the reaction, \cee{ A + D  <=> DA }, where A and D, represent acceptor and donor, water molecules. Using the equilibrium constant computed at different temperatures, we obtain $\Delta U =$ -18.68 $\pm$ 2.05 kJ/mol and $\Delta S$ = -24.79 $\pm$ 5.83 J/K mol. These values are only marginally affected by small changes in the geometric criteria used to define a hydrogen-bond (see SI for details). While hydrogen-bonding is just one of many factors contributing to the total entropy of an aqueous system, the high concentration of water in liquid water (55 mol/L) means that even small changes in the fraction of hydrogen-bonded molecules can have a significant thermodynamic impact. In this context, our results show that, prior to the onset of dissociation, the applied electric field drives the system toward a highly ordered, low-entropy state.

\subsection*{Proton defects disrupt hydrogen-bond network}

\begin{figure*}
\centering
\includegraphics[width=0.9\columnwidth]{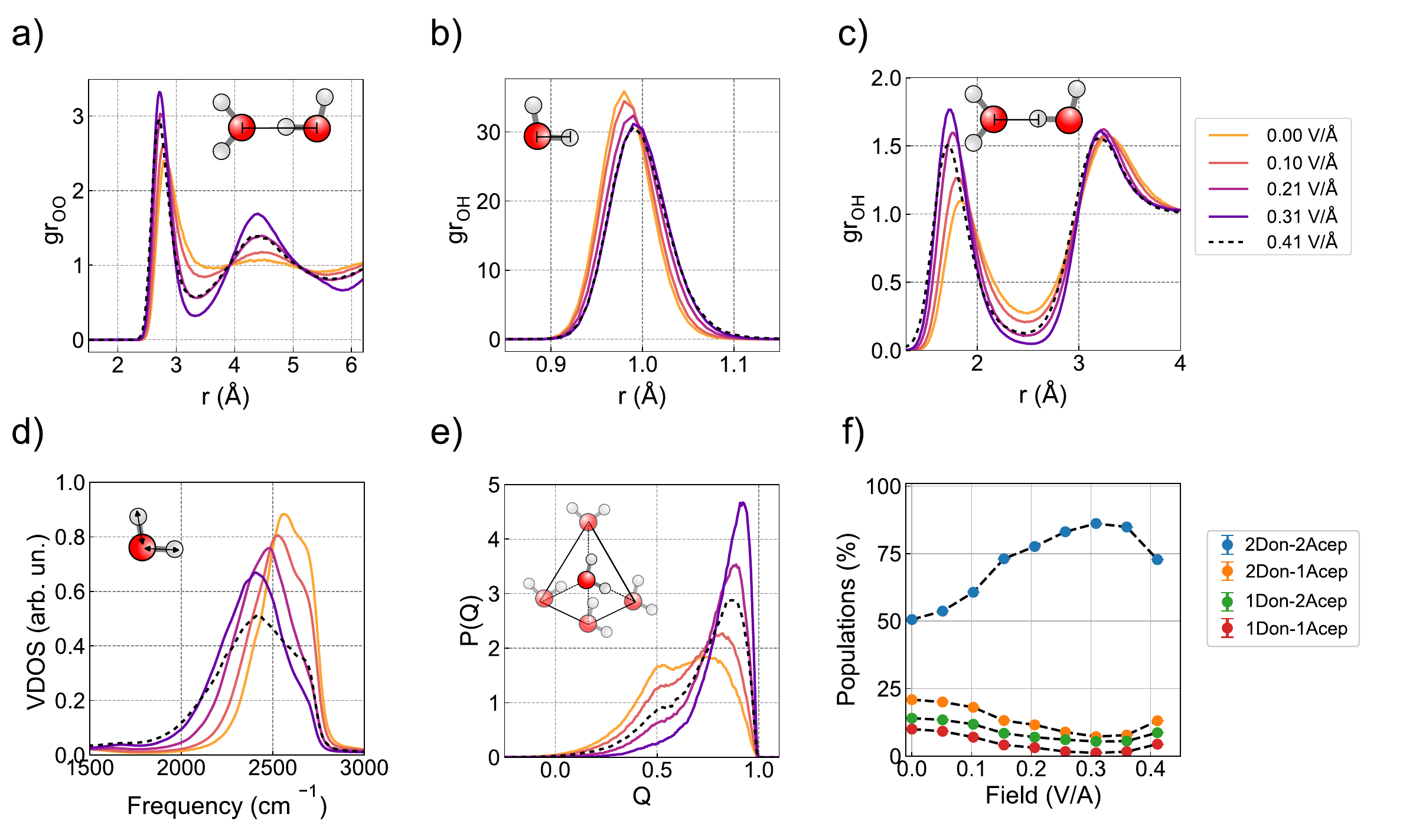}
\caption{\textbf{Water and hydrogen-bond network structure under electric fields.} a-c) Oxygen-oxygen and oxygen-hydrogen radial distribution functions, 
d) vibrational density of states (VDOS) of water stretching mode, e)
tetrahedral order parameter, and f) classification of water molecules according to the hydrogen bonding environment. 
Note that the main peak of the VDOS in panel d appears at approximately 2500 cm$^{-1}$ due to the use of deuterium masses for hydrogen atoms.
}
\label{fig:structure}
\end{figure*}

At field strengths exceeding $\textbf{E}=0.36$ V/\AA, the concentration of ions surpasses 100 mM, significantly disrupting the hydrogen-bond network  (dashed lines in Fig. \ref{fig:structure}a-e), and therefore reversing the trends observed at lower field. Specifically, the height of the first peak of the gr$_\text{OO}$ and the second peak of the gr$_\text{OH}$ decreases, the $Q$ distribution shows a lower tetrahedral order, and the number of 2Don-2Acep decreases. Together, these changes indicate that the formation of proton and hydroxide species weakens the overall hydrogen-bond network.

\begin{figure}
\centering
\includegraphics[width=0.3\columnwidth]{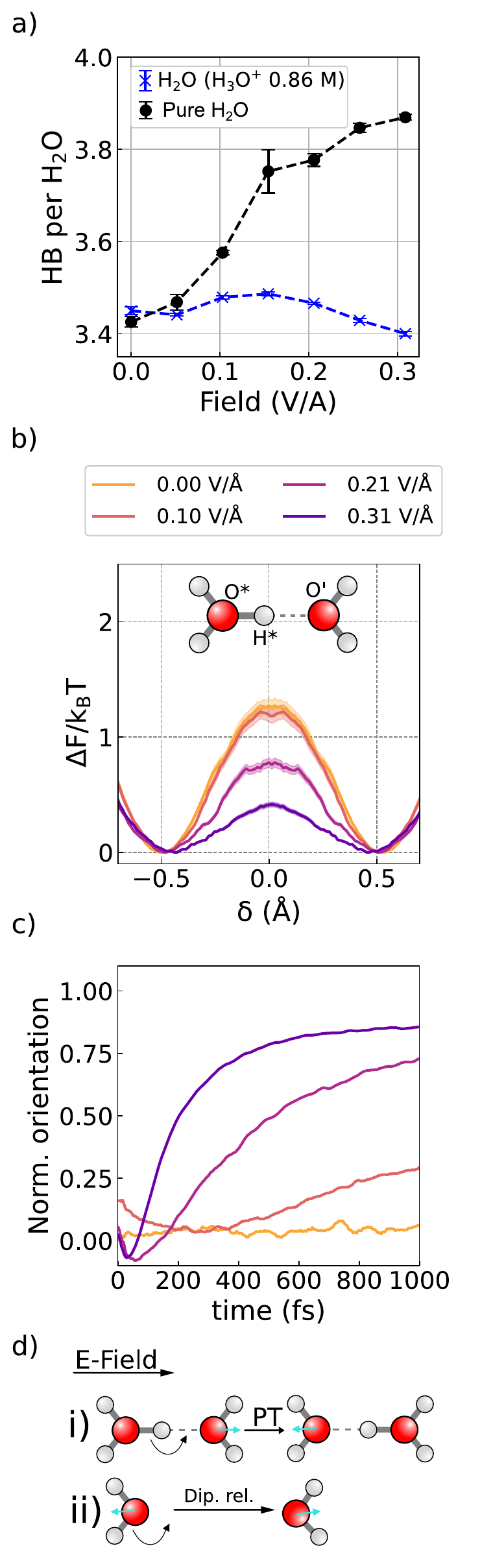}
\caption{\textbf{Proton conduction under electric fields disrupts the hydrogen-bond network.} a) Number of hydrogen-bonds (HB) per water molecule for pure water and 0.86 M aqueous proton solution. b) Proton transfer free energy barrier along the proton sharing coordinate, $\delta$ (see main text for its definition). c) Average orientation of newly created water molecules after a proton transfer event. d)  Schematic representation of proton-induced water reorientation (i) Proton transfer (PT) event leads to a newly formed water molecule, which (ii) reorientates its dipole to align it parallel to the external electric field (dipole relaxation). During this reorientation, existing hydrogen bonds are broken and new ones are formed. }
\label{fig:ions}
\end{figure}

To confirm the disruptive effect of ions on water structuring at finite electric fields, we performed additional AIMD simulations on systems containing a permanent proton or hydroxide ion.
Similar results were obtained for both types of systems, so for clarity, we discuss only the results for systems with a permanent proton.
The field-induced water alignment with the external field closely resembles that of pure water, indicating that the presence of ions has only a minor effect on the dielectric saturation (see Fig. S8 in the SI). 
However, when monitoring the average number of hydrogen-bonds, a stark difference emerges. As shown in Fig. \ref{fig:ions}a, in pure water, the number of hydrogen-bonds steadily increases with field strength, consistent with the structural trends discussed earlier. In contrast, in the system containing an additional proton (blue curve in Fig. \ref{fig:ions}a), the number of hydrogen-bonds remains nearly constant across all field strengths. These results demonstrate that the presence of protons strongly inhibits the field-induced structuring of the hydrogen-bond network.

To understand this disruptive effect, we calculated the proton transport free energy barrier. For this, we used the standard definition of the proton-sharing coordinate, 
$\delta = d_\text{O*H*} - d_\text{O'H*}$, where $d$ denotes the distance between the respective atoms,  and * and ' refer to atoms belonging to the proton defect and nearest water molecule, respectively. For each frame, H* was selected to obtain the lowest of the possible $\delta$. In the absence of an external field, the computed proton transfer barrier is  1.2 k$_\text{B}$T, in reasonable agreement with previous AIMD simulations \cite{Atsango_JCP_2023,Advincula_ACSNano_2025}.
At finite field strengths, the barrier decreases with increasing field, suggesting a faster proton diffusion, as expected for a charged particle subjected to an electric field.

Although a faster proton diffusion explains the increased mobility, it alone is insufficient to account for the disruption of the hydrogen-bond network.  Proton transfer events require a specific geometric arrangement of the hydronium ion, H$_3$O$^+$, and an acceptor water molecule. In Fig. \ref{fig:ions}c, we present the average orientation of newly created water molecules, where we define time zero as the moment of the proton hop. Immediately after the proton transfer, the newly created water molecule exhibits no net orientation. Over the next few hundred femtoseconds, it reorients to align with the external field, a process that induces the breaking of its preexisting hydrogen-bonds. Thus, the field-enhanced proton transfer acts as a local, repeated disruption mechanism, progressively weakening the hydrogen-bond network and driving the system toward a more disordered, higher-entropy state (see schematic diagram in Fig. \ref{fig:ions}d). This disruptive phenomenon was equally observed for simulations containing a permanent hydroxide ion (see Fig. S9 in the SI). 

\subsection*{Discussion}

In this work, we presented  AIMD simulations of bulk water under varying external electric fields and temperatures. Consistent with previous studies, we found that external fields act as structure-makers at low intensities and induce water autoionization at higher fields. We computed the reaction free energy, reaction internal energy ($\Delta U$), and reaction entropy  ($\Delta S$) associated with the WD reaction, showing that the reaction is entropically hindered in the absence of a field but becomes entropy-driven under finite fields.
Furthermore, through additional simulations involving proton and hydroxide ions, we demonstrated that proton defects inhibit the field-induced water structuring by randomizing molecular orientations during proton transfer events, thereby weakening the hydrogen-bond network.

Putting together all these results, we are now in a position to rationalize the sign of the WD reaction entropy at large fields. This can be understood by comparing the thermodynamic states of the reactants (pure water) and the products (aqueous proton and hydroxide). At high fields, the reactant state becomes highly ordered: water molecules are strongly aligned with the field, the number of hydrogen-bonds increases, and the hydrogen-bonds are strengthened. Overall, the system is driven into a highly structured, low-entropy state.
By contrast, the product state, due to the presence of disruptive ionic defects, exhibits a reduced orientational order, a lower number of hydrogen-bonds, and weaker bonding. These changes contribute to an increase in rotational entropy. In addition, translational entropy rises, as one water molecule produces two highly mobile ionic species.
The combined effect of these contributions gives rise to the large positive reaction entropy observed at high fields.
A precise quantification of the individual entropic components remains challenging and is still an active area of research \cite{Peter_JCP_2004}. In this context, Cassone \textit{et al.} recently estimated rotational and translational entropy changes using classical and \textit{ab initio} simulations via the two-phase thermodynamics (2PT) formalism \cite{Nibali_JCP_2023}. Their study found relatively small entropy reductions (a few J/K mol), although only field strengths below the dissociation threshold were considered. Their findings are consistent with our results and highlight the crucial role of ionic defects in altering the field-induced response of water.

The field-induced disruption mechanism proposed here resembles that suggested to explain the lower proton conductivity of ferroelectric ice XI compared to orientationally disordered ice I$\textrm{h}$ \cite{Cassone_JPCB_2014}. In short, the reduced conductivity of ice XI arises from the difficulty in restoring dipole alignment due to the generation of proton traps \cite{Park_JPCB_2014,Hassanali_CPL_2014}: after each proton transfer event, the entire network of molecular dipoles must reorient to enable further transfers along a given path.
At the same time, ice XI exhibits a lower onset field for WD than ice  I$\textrm{h}$. We hypothesize that such a greater propensity for the WD reaction stems from similar entropic effects to those described above.

Let us now turn to the reaction energy. 
Simple electrostatic arguments would predict that an external electric field stabilizes charge separation, thereby lowering $\Delta U$. However, our results show that this is not the case. Instead, the molecular nature of the proton and hydroxide leads to a more complex scenario, in which compensating effects arise. Specifically, the same weakening and reduction of hydrogen-bonds that contribute to the positive $\Delta S$ also introduce a positive enthalpic contribution. This offsets the negative enthalpic contribution expected from charge stabilization, resulting in an overall relatively modest change in $\Delta U$ with increasing field strength. We believe that this competition may underlie the non-monotonic dependence of $\Delta U$  reported in Table \ref{tab:Free_energy}.

Proton defects are well known to form strong hydrogen-bonds in water, leading to vibrational frequency shifts of several hundred cm$^{-1}$, which arise in part from the strongly coupled vibrations of multiple water molecules solvating the defects~\cite{Mandal_JCP_2014,Fournier_NatChem_2018,Agmon_ChemRev_2016}. Recently, Car and co-workers demonstrated that structural correlations between proton defects and neighboring water molecules exhibit sharper peaks than correlations between water molecules themselves, indicating that the local environment around proton defects is, on average, more ordered \cite{Andrade_PNAS_2023}. 
These findings are in agreement with the experimentally measured negative partial molar volumes for protons and hydroxide ions, confirming a more structured and compressed hydrogen-bond network near these ions \cite{Borsarelli_JPPB_1998,Marcus_JCP_2012}.
Moreover, the WD reaction itself is associated with a large negative reaction entropy, mainly originating from intermolecular interactions, further supporting the notion that proton defects act as structure makers under standard conditions.
In contrast, our results show that under finite electric fields, proton and hydroxide ions behave as structure breakers. This demonstrates that an external field can modulate and qualitatively transform ion–solvent interactions, turning classical structure makers into structure breakers. Future work will investigate whether this field-induced change in behavior also extends to other inorganic 'spectator' ions.

Our AIMD simulations are based on DFT employing the revPBE-D3 exchange correlation functional.
This functional belongs to the generalized gradient approximation (GGA) family and is known to overestimate hydrogen-bond strengths and underestimate hydrogen transfer barriers  \cite{Atsango_JCP_2023,Litman_FardayDiscuss_2020, Dasgupta_JPCL_2025,Gillan_JCP_2016}. Additionally, while the inclusion of nuclear quantum effects (NQEs) further modulates the strength of the hydrogen-bond network and lowers the field-induced dissociation threshold~\cite{Cassone_JPCL_2020}, it is well-known that combining NQEs with a GGA-level potential energy surface leads to a drastic overestimation of these effects~\cite{Atsango_JCP_2023,Litman_FardayDiscuss_2020,Ceriotti_ChemRev_2016}. Thus, omitting NQEs at the GGA level often results in better accuracy due to a partial cancellation of errors. Indeed, this likely explains why our simulations yield a surprisingly accurate value of pKw (see Table \ref{tab:Free_energy}). To go beyond these limitations, the use of machine learning interatomic potentials (MLIPs) is required.
Recent developments in that area have enabled large-scale simulations using hybrid and meta-GGA functionals~\cite{Andrade_PNAS_2023,Dasgupta_JPCL_2025}, as 
well as explicitly correlated methods \cite{Oneill_JPCL_2024,Chen_JCTC_2023,Daru_PRL_2022}, and new approaches that allow coupling to external fields have been proposed \cite{SCFNN,Joll_NatCom_2024,stocco2025,Jana_JCIM_2024}. However, in most cases, the architecture is based on zero-field expansion or relies on a reference frame,  limiting their ability to capture water dissociation under strong fields. Our group is actively developing next-generation MLIPs that will allow the accurate simulation of field-driven dissociation processes and enable the study of more complex systems with higher accuracy and longer timescales.

Despite the simplicity of our system, the decisive role of field-induced solvent ordering observed for the WD reaction could have important implications in other contexts, such as the hydrogen evolution reaction (HER), where rates depend sensitively on the ordering of interfacial water molecules under electric fields \cite{Li_NatCat_2022,Wang_Nature_2021}.
In most cases, such rate enhancements are rationalized in terms of purely energetic stabilization of transition-state geometries.
However, recent studies suggest that increased activation entropy due to proton delocalization could govern HER rates in certain materials \cite{Chen_Joule_2023,Rodellar_NatEner_2024,Gisbert-González_JACS_2025}.
Our simulations provide strong evidence for such field-induced proton delocalization and, more broadly, support the emerging view that field-induced molecular organization and solvent alignment, namely entropic effects, may be central to accelerate numerous reactions at aqueous interfaces~\cite{Nam_PNAS_2017,Chamberlayne_JCP_2020,Ruiz-Lopez_NatRev_2020},  highlighting the need for more temperature-dependent studies of interfacial aqueous reactions.

In conclusion, our work reveals the underlying thermodynamic factors that govern arguably the simplest field-induced aqueous reaction, the  WD reaction. Given the parallels with processes relevant to atmospheric chemistry, electrochemistry, and biochemistry, we propose that entropic effects are likely more ubiquitous and influential than previously recognized. These insights could prove crucial in the rational design of novel aqueous catalysts for "on-water" and electrochemical reactions.

\subsection*{Methods}

\textbf{System Setup} 

The pure water system consisted of 64 H$_2$O molecules in a cubic box with periodic boundary conditions and a side length of 12.42 Å. Aqueous proton (hydroxide) systems were generated by replacing a water molecule with a single H$_3$O$^+$ (or OH$^-$) in the same setup, resulting in 63 H$_2$O + 1 H$_3$O$^+$ (or 63 H$_2$O + 1 OH$^-$).
In the SI, we provide the results of additional simulations with 128 H$_2$O molecules (and 127 H$_2$O + 1 H$_3$O$^+$ / 1 OH$^-$) in a box with a side length of 15.644 Å to assess finite-size effects.

\textbf{AIMD}
AIMD simulations were performed with the CP2K code \cite{kuhne_cp2k_2020}, using the revPBE exchange–correlation functional \cite{revPBE} augmented with D3 dispersion corrections \cite{D3}. Kohn–Sham orbitals were expanded in a TZV2P basis set, and the electron density was represented with an auxiliary plane-wave basis with a cutoff of 400 Ry. This setup has been shown to reproduce the structural and dynamical properties of water reasonably well \cite{Marsalek_JPCL_2017}.
Hydrogen atoms were assigned a mass of 2.0 amu, and a timestep of 2.0 fs was used to integrate the equations of motion. Simulations were carried out in the NVT ensemble using the stochastic velocity rescaling thermostat \cite{Bussi_JCP_2007} with a 50 fs time constant, discarding the first 10 ps for thermalization. Unless otherwise stated, trajectories were run for 100 ps after equilibration.
Error bars were computed by dividing the trajectory into ten 10-ps blocks and calculating the standard deviation of the mean.
To obtain $\Delta S$ and $\Delta U$ simulations were run at  300 K, 330K, 360K, and 390 K. We note that the highest temperature corresponds to a superheated regime of water. Simulations under external electric fields were initialized from pre-thermalized structures at lower field strengths. To ensure numerical stability, the field was increased incrementally in steps of 0.001 Hartree/Bohr ($\sim$0.05 V/Å).

The tetrahedral order parameter of the $i$-th water molecule, $Q_i$, was computed as \cite{Errington_Nature_2001},

        \begin{equation}
     Q_i  =  1 - \frac{3}{8}\sum_{j=1}^3\sum_{k=j+1}^4(\cos \Psi_{ijk}+\frac{1}{3})^2
       \end{equation}
       \nid where $\Psi_{ijk}$ is the angle formed by the molecule i and its nearest neighbors j and k.

\textbf{Enhanced Sampling Simulations}

Free energy calculations of the WD reaction at $\textbf{E} = 0.00$ V/Å and $\textbf{E} = 0.18$ V/Å required enhanced sampling techniques. We employed umbrella integration \cite{UI}, which combines thermodynamic integration and umbrella sampling to reduce statistical errors. Simulations were performed using i-PI \cite{ipiv3} coupled to CP2K \cite{kuhne_cp2k_2020} and PLUMED \cite{plumed}.
As a collective variable, we used the coordination number of a selected oxygen atom, defined as:

\begin{equation} n_\text{cov} = \sum_{i \in H} \frac{1-\left(\frac{r_{i}}{R_0}\right)^{12}}{1-\left(\frac{r_{i}}{R_0}\right)^{24}},
\end{equation}

\nid where the sum runs over all hydrogen atoms, $R_0 = 1.38$ Å, and $r_{i}$ is the distance between the selected oxygen atom and the $i$-th hydrogen.
Umbrella integration used 11 windows spanning $n_\text{cov}$ values from 2.0 to 1.0, with a harmonic bias of 200 kcal/mol~\cite{Joutsuka_JPCB_2022,Dasgupta_JPCL_2025}.
We used $n_\text{cov}=2.0$ and $n_\text{cov}=1.2$, to define the intact water and hydroxide states, respectively~\cite{DeLaPuente_JACS_2023}.
Although the need for a second collective variable \cite{Joutsuka_JPCB_2022, Dasgupta_JPCL_2025} or more sophisticated ones based on Voronoi tesselation \cite{Grifoni_PNAS_2019,Andrade_PNAS_2023} has been suggested, the limited simulation times in AIMD mean that these refinements would be inconsequential relative to our statistical error bars.

\begin{acknowledgments}
Y.L. would like to thank the members of the ICE group, S. Brookes, X. R. Advincula, F. Kanoufi, D. Scherlis, M. Lizee, and S. Öner for stimulating discussions, and G. Cassone for sharing the CP2K input files employed in Ref. \cite{Cassone_JPCL_2020}.
Y.L. and A.M. acknowledge support from the European Union under the "n-AQUA" European Research Council project (Grant No. 101071937).
We further acknowledge computational resources provided by the UK national high-performance computing service, ARCHER2, accessed via the UKCP consortium and funded by EPSRC grant EP/X035891/1. 
\end{acknowledgments}

\textbf{Data availability} All data required to reproduce the findings of this study will be made available upon acceptance of this manuscript in a public repository.

\end{document}


\preprint{AIP/123-QED}

\title{Supporting information: Entropy governs the structure and reactivity of water dissociation under electric fields}
\author{Yair Litman}
 \affiliation{Yusuf Hamied Department of Chemistry,  University of Cambridge,  Lensfield Road,  Cambridge,  CB2 1EW, UK}
 \affiliation{Max
Planck Institute for Polymer Research, Ackermannweg 10, 55128 Mainz,
Germany}
 \email{litmany@mpip-mainz.mpg.de}
  \email{ am452@cam.ac.uk}
\author{Angelos Michaelides}
 \affiliation{Yusuf Hamied Department of Chemistry,  University of Cambridge,  Lensfield Road,  Cambridge,  CB2 1EW, UK}

\date{\today}

\maketitle

\subsection*{Additional convergence and robustness tests}
\begin{figure}
\centering
\includegraphics[width=0.7\textwidth]{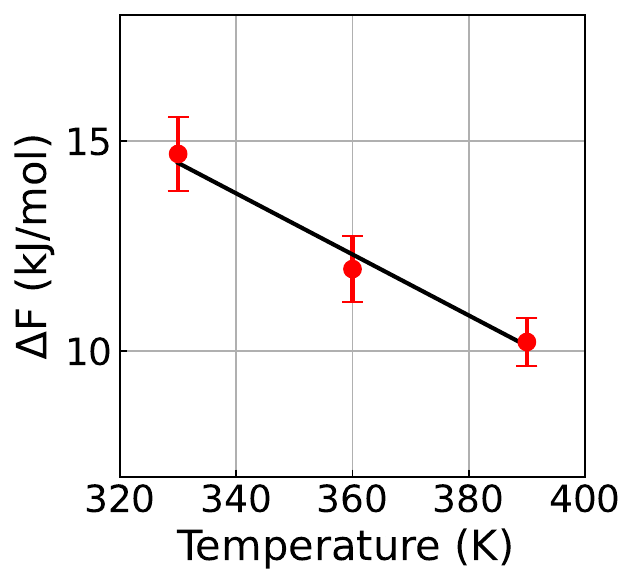}
\caption{ Helmholtz free energy of the water dissociation reaction as a function of temperature obtained from AIMD simulations with BLYP-D3 exchange correlation functional at \textbf{E}=0.36 V/\AA. Linear fit of the data leads to $\Delta U$= 38.44 $\pm$ 6.28 kJ/mol and $\Delta$S=72.61 $\pm$ 16.99 J/mol K.}
\label{fig:BLYP}
\end{figure}

~\\
To assess the sensitivity of our results to the specific choice of exchange–correlation functional, we repeated the calculation using a different GGA functional. 
In Fig.~\ref{fig:BLYP}, we present the free energy of the water dissociation reaction at different temperatures at \textbf{E} = 0.36 V/Å obtained with the BLYP-D3 functional. The fitted values of $\Delta U$ and $\Delta S$ are 20–25\% smaller in magnitude compared to those obtained with revPBE-D3 presented in the main text. Importantly, in both cases, $\Delta S$ remains large and positive, demonstrating that our conclusions are robust within this level of theory.  Because classical nuclei simulations using GGA functionals are known to be benefited by the partial error cancelation arising from the neglect of nuclear quantum effects and the underestimation of the hydrogen transfer barrier \cite{Ceriotti_ChemRev_2016, Atsango_JCP_2023}, we do not anticipate qualitative changes if more accurate simulations are performed, for example employing hybrid or meta-GGA functionals and including nuclear quantum effects.

\begin{figure}
\centering
\includegraphics[width=0.7\textwidth]{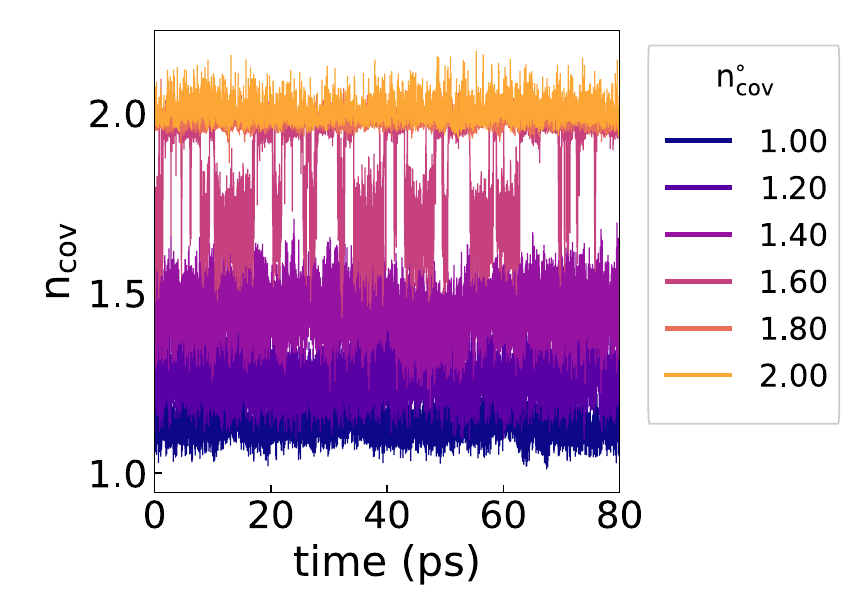}
\caption{Time traces of selected umbrella windows obtained from revPBE-D3 simulations at 330 K in the absence of an external field.
$n_\text{cov}^\circ$  corresponds to the coordination number value set for each umbrella window, and $n_\text{cov}$ represents the instantaneous coordination number. For clarity, only every second umbrella window is shown.}
\label{fig:Umbrella}
\end{figure}

Fig. \ref{fig:Umbrella} shows the time traces of the umbrella simulations for a selected system. It can be observed that the reaction coordinate is adequately sampled between the reactant and product states.

\begin{figure}
\centering
\includegraphics[width=0.4\textwidth]{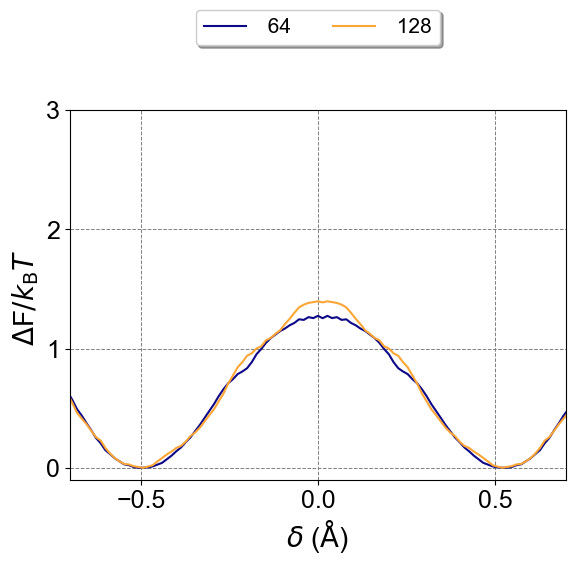}
\caption{ Proton transfer free energy along the proton sharing coordinate for a system containing 
64 oxygen atoms (63 H$_2$O + 1 H$_3$O$^+$)  and 128 oxygen atoms (127 H$_2$O + 1 H$_3$O$^+$)}
\label{fig:proton1}
\end{figure}

\begin{figure}
\centering
\includegraphics[width=0.4\textwidth]{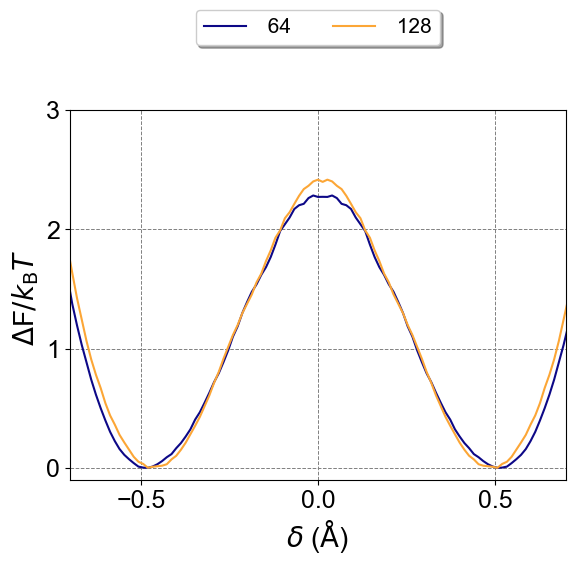}
\caption{ Proton transfer free energy along the proton sharing coordinate for a system containing 64 oxygen atoms (63 H$_2$O + 1 OH$^-$)  and 128 oxygen atoms (127 H$_2$O + 1 OH$^-$).}
\label{fig:hydroxide}
\end{figure}

\begin{figure}
\centering
\includegraphics[width=0.7\textwidth]{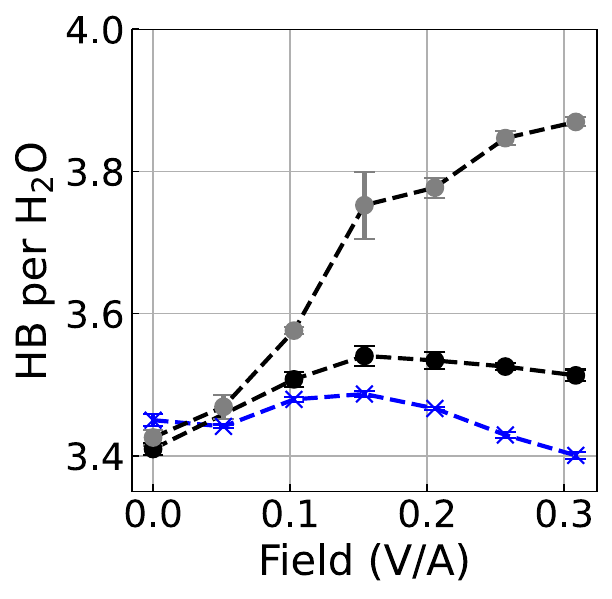}
\caption{ Number of hydrogen-bonds (HB) per water molecule for pure water (gray circles)  and aqueous proton solution for boxes containing 64 oxygen atoms, 63 H$_2$O + 1 H$_3$O$^+$,  and 128 oxygen atoms, 127 H$_2$O + 1 H$_3$O$^+$ (blue crosses and black circles respectively) }
\label{fig:proton2}
\end{figure}

To verify the small impact of finite size effects reported in Ref. \cite{Cassone_JPCL_2020}, we repeated some simulations, doubling the volume of the simulation box. We used a 15.644 \AA$^3$ simulation box containing 127 H$_2$O molecules and 1 H$_3$O$^+$ or 1 OH$^-$.
Fig. \ref{fig:proton1} and \ref{fig:hydroxide} show the proton transfer free energy profiles for systems containing a permanent proton and hydroxide ion, respectively. In both cases, simulations with 64 and 128 oxygen atoms yield comparable energy profiles. Fig. \ref{fig:proton2}, shows the average number of hydrogen bonds per water molecule at different external field strengths for pure water and aqueous proton in the small and large box sizes. In both scenarios, the proton exhibits a pronounced and comparable structure‐breaking effect.

\begin{figure}
\centering
\includegraphics[width=0.7\textwidth]{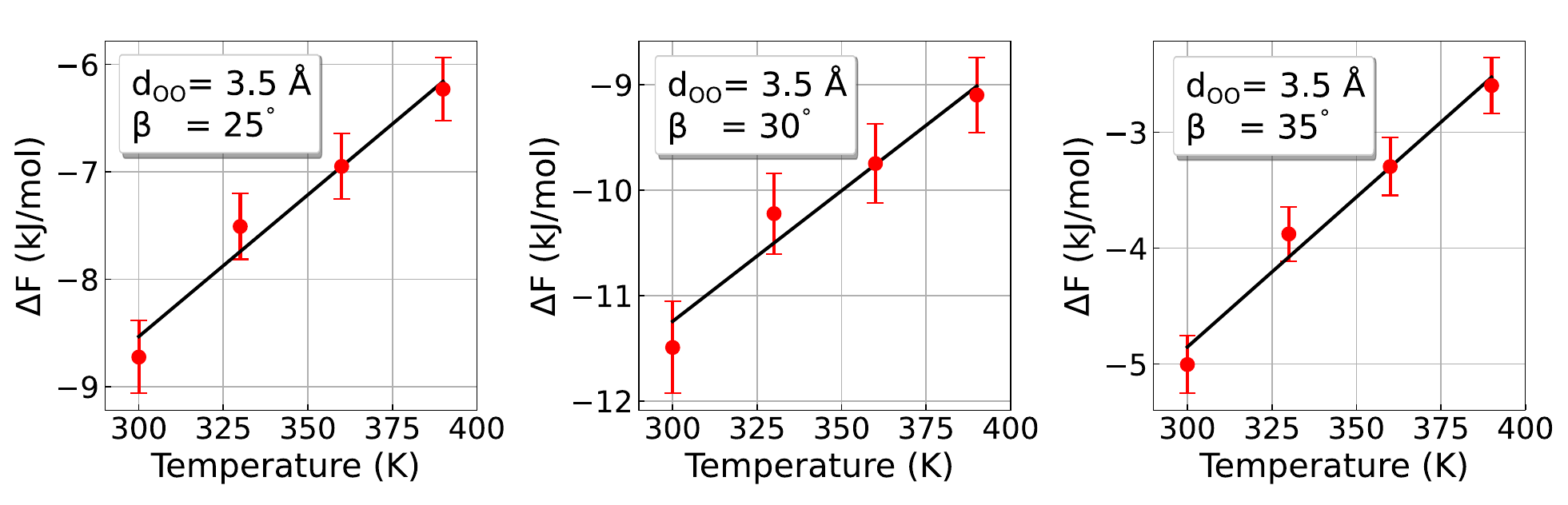}
\caption{ Temperature dependence of the hydrogen-bond formation free energy for different angle definitions }
\label{fig:hydrogen_bond_angle}
\end{figure}

\begin{figure}
\centering
\includegraphics[width=0.7\textwidth]{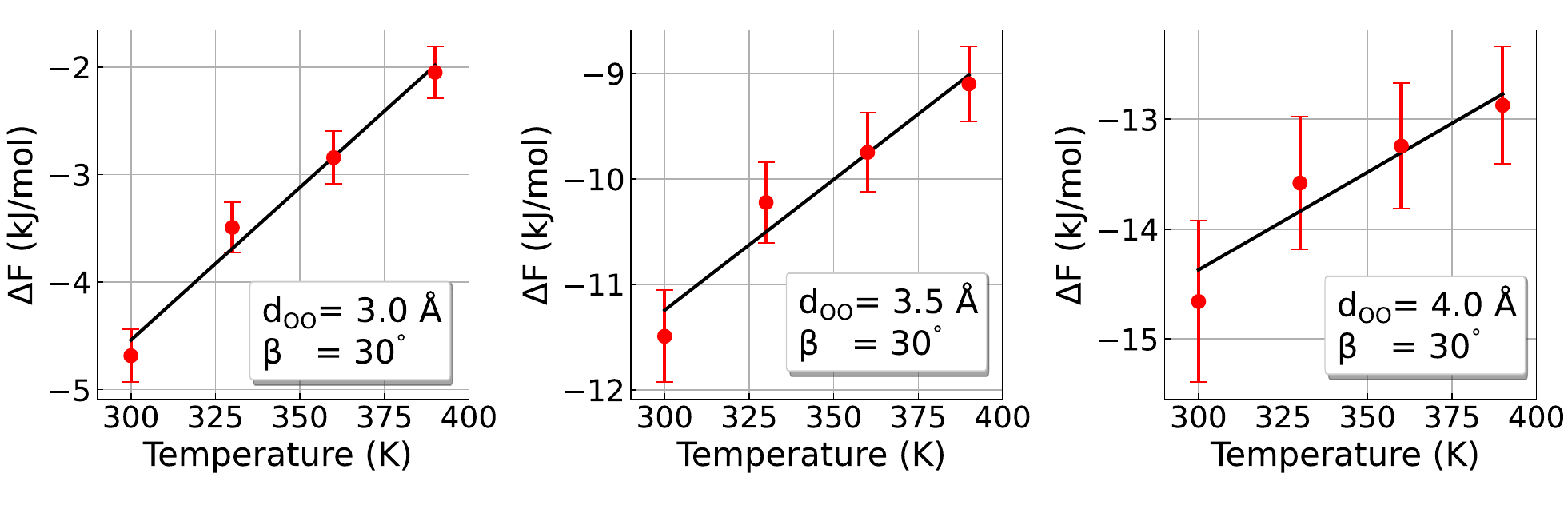}
\caption{ Temperature dependence of the hydrogen-bond formation free energy for different distance definitions. Fitted values for $\Delta U$ and $\Delta S$ are reported in Tab. \ref{tab:hb_definition}.  Fitted values for $\Delta U$ and $\Delta S$ are reported in Tab. \ref{tab:hb_definition}. }
\label{fig:hydrogen_bond_distance}
\end{figure}

Fig. \ref{fig:hydrogen_bond_angle} and \ref{fig:hydrogen_bond_distance} show the temperature dependence of the free energy associated with hydrogen bond formation, using slightly different angle and distance criteria, respectively. Tab. \ref{tab:hb_definition} presents the corresponding formation energies and entropies, showing that variations in the hydrogen bond definition have only a minimal impact on its thermodynamics.

\begin{figure}
\centering
\includegraphics[width=0.7\textwidth]{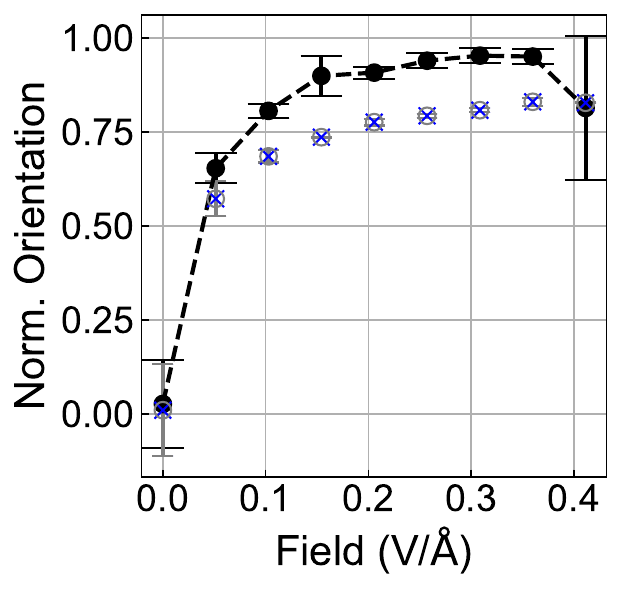}
\caption{Average orientation of molecular dipoles along the field
direction for pure water (black filled circles), aqueous proton (blue crosses), and
aqueous hydroxide (gray empty circles). A value of 1.00 represents a perfect
alignment between the molecular dipoles and the external field.}
\label{fig:orientation_with_ions}
\end{figure}

Fig. \ref{fig:orientation_with_ions} shows the field-induced water alignment for pure water and acid and basic solutions, where it is direct to observe that the presence of ions has only a minor effect on the dielectric saturation.

\begin{figure*}
\centering
\includegraphics[width=.95\textwidth]{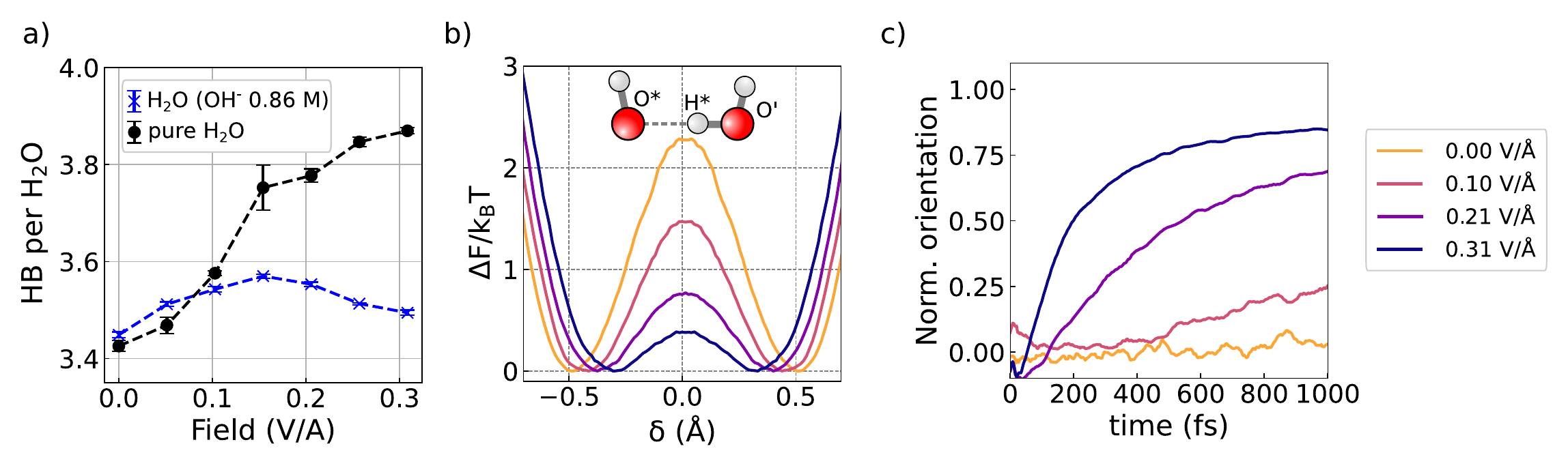}

\caption{a) Number of hydrogen-bonds (HB) per water molecule for pure water and 0.86 M aqueous hydroxide solution. b) Proton transfer free energy barrier along the proton sharing
coordinate, $\delta$ (see main text for its definition). c) Dipole relaxation of the newly formed water molecule after a proton transfer event.}
\label{fig:ions_OH}
\end{figure*}

Fig. \ref{fig:ions_OH} presents the number of hydrogen bonds, the proton transfer free energy barrier, and the relaxation dynamics of newly formed water molecules for a system containing an OH$^-$. Similar to that reported for H$_3$O$^+$ in the main text, OH$^-$ exhibits a markedly structure-breaking behavior.

\begin{table}[h!]
\centering
\begin{tabular}{|c|c|c|c|}
\hline
d$_\text{OO}$ &$\beta$ & $\Delta U$ (kJ/mol)& $\Delta S$ (J/ K mol) \\ \hline
\hline
 30.00 &  3.00 & -13.05 $\pm$  1.25 & -28.38 $\pm$   3.61\\
 30.00 &  3.50 &-18.68 $\pm$   2.05 & -24.79 $\pm$   5.83\\
 30.00 &  4.00 &-19.71 $\pm$   3.28 & -17.78 $\pm$   9.29\\
 \hline
 25.00 &  3.50 &-16.45 $\pm$   1.64 & -26.38  $\pm$  4.68\\
 30.00 &  3.50 &-18.68 $\pm$   2.05 & -24.79  $\pm$  5.83\\
 35.00 &  3.50 &-21.80 $\pm$   3.11 & -23.92  $\pm$  8.77\\
\hline
\end{tabular}
\caption{ Formation energy and entropy of hydrogen bond using different distance, d$_\text{OO}$, and angle, $\beta$, thresholds.  }
\label{tab:hb_definition}
\end{table}

